\documentstyle[11pt]{article}

\begin{document}

\author{John D. Barrow and Kerstin E. Kunze \\
Astronomy Centre\\
University of Sussex\\
Brighton BN1 9QH\\
U.K.}
\title{Spatially Homogeneous String Cosmologies}
\maketitle
\date{}

\begin{abstract}
We determine the most general form of the antisymmetric $H$-field tensor
derived from a purely time-dependent potential
that is admitted by all possible spatially homogeneous cosmological models
in 3+1-dimensional low-energy bosonic string theory. The maximum number of
components of the $H$ field that are left arbitrary is found for each
homogeneous cosmology defined by the Bianchi group classification. The
relative generality of these string cosmologies is found by counting the
number of independent pieces of Cauchy data needed to specify the general
solution of Einstein's equations. The hierarchy of generality differs
significantly from that characteristic of vacuum and perfect-fluid
cosmologies. The degree of generality of homogeneous string cosmologies is
compared to that of the generic inhomogenous solutions of the string field
equations.\\

\end{abstract}

\begin{center}
{\bf I. INTRODUCTION}
\end{center}

The low-energy effective action of the bosonic sector of string theory
provides cosmological models that might be applicable just below the Planck
(or string) energy scale in the very early universe [1]. A number of studies
have been made of these cosmologies in order to ascertain the behaviour of
simple isotropic and anisotropic universes, investigate the implications of
duality, and search for inflationary solutions [2-6]. Many of the
traditional questions of general relativistic cosmology can be asked of the
cosmological models defined by string theory: do they possess space-time
singularities?, what is the generic behaviour of the solutions at late and
early times?, what exact solutions can be found in closed form?, and what
relation do particular exact solutions have to the general cosmological
solution? Since this theory is to be applied at times very close to the
Planck epoch it would be unwise to make special assumptions about the form
of the cosmological solutions. Anisotropies and inhomogeneities could play
an important role in the evolution. Indeed, any dimensional reduction
process could be viewed as an extreme form of anisotropic evolution in D
dimensions in which three spatial dimensions expand whilst the rest remain
static. Because of these irreducible uncertainties about the very early
Universe, one would like to understand the general behaviour of wide classes
of solution in order to ascertain the relative generality of any particular
solution that may be found. A number of studies have focused on obtaining
particular solutions for 3+1 dimensional space-times in cases where spatial
homogeneity (and sometimes also isotropy) is assumed for the metric of
space-time, where the $H$ field is set to zero [4], or where the $H$ field
is included by assuming that it takes a particular form which satisfies its
constraints and its equation of motion [5]. For example, Copeland et al,
[2], discussed Friedmann and Bianchi type I universes, allowing $*H$ to be
time-dependent or space-dependent, respectively. In a second paper, [3],
they discussed Bianchi I solutions with a homogeneous antisymmetric tensor
field. 
In [6] (see also [5]) Batakis presented an overview of all possible
configurations of a (spatially) homogeneous $H$-field in diagonal
Bianchi models with a metric
$$ds^{2}=-dt^{2}+a_{1}(t)^{2}(\omega^{1})^{2}+a_{2}(t)^{2}
(\omega^{2})^{2}+a_{3}(t)^{2}(\omega^{3})^{2}$$
where $\{dt,\omega^{\alpha}\}$ is the standard basis.
However, in this paper the Bianchi models are not assumed to be
diagonal. 

The form of the $H$-field derived from a time-dependent potential will
be determined in all 
four-dimensional space-times with homogeneous three-spaces. These
three-spaces were first classified by Bianchi [7] and have been extensively
studied in the cosmological context following their introduction into
cosmology by Taub [8]. They provide us with the general class of
cosmological models whose solutions are determined by ordinary differential
equations in time. By generalising a procedure used to study electromagnetic
fields in spatially homogeneous cosmological models by Hughston and Jacobs
[10], we can determine the maximum number of components permitted for the $H$
field in each of the Bianchi cosmologies. This enables us to determine the
number of degrees of freedom which define the string cosmology of each case.
The results are interesting. The Bianchi types containing the most general
geometries place the most restrictions upon the presence of the $H$ field.

The string world sheet action for a closed bosonic string in a background
field including all the massless states of the string as part of the
background is given by, [1], 
\begin{equation}
S=-\frac 1{4\pi \alpha ^{\prime }}\int d^2\sigma \{\sqrt{h}h^{\alpha \beta
}\partial _\alpha X^\mu \partial _\beta X^\nu g_{\mu \nu }(X^\rho )+\epsilon
^{\alpha \beta }\partial _\alpha X^\mu \partial _\beta X^\nu B_{\mu \nu
}(X^\rho )+\alpha ^{\prime }\sqrt{h}\phi (X^\rho )R^{(2)}\}
\end{equation}
where $h^{\alpha \beta }$ is the 2-dimensional worldsheet metric, $R^{(2)}$
the worldsheet Ricci scalar, $\epsilon ^{\alpha \beta }$ the worldsheet
antisymmetric tensor, $B_{\mu \nu }(X^\rho )$ the antisymmetric tensor
field, $g_{\mu \nu }(X^\rho )$ the background space-time metric (graviton), $%
\phi (X^\rho )$ the dilaton, $\alpha ^{\prime }$ is the inverse string
tension, and the functions $X^\rho (\sigma )$ map the string worldsheet into
the physical D-dimensional space-time manifold.

For the consistency of string theory it is essential that local scale
invariance holds. Imposing this condition results in equations of motion for
the fields $g_{\mu \nu }$, $B_{\mu \nu }$ and $\phi $ which can be derived
to lowest order in $\alpha ^{\prime }$ from the low-energy effective action 
\begin{equation}
S=\int d^Dx\sqrt{-g}e^{-\phi }(R+g^{ab}\partial _a\phi \partial _b\phi
-\frac 1{12}H^{abc}H_{abc}-\Lambda).
\end{equation}

In this paper we assume a vanishing cosmological constant, $\Lambda$.

In a cosmological context it is generally assumed that by some means 
all but four of the 10 or 26 dimensions of space-time are compactified,
leaving an expanding 3+1-dimensional space-time ($D=4)$. Since we are
interested in cosmological solutions of the field equations derived from the
variation of this action, we adopt the Einstein frame by making the
conformal transformation 
\begin{equation}
g_{ab}\rightarrow e^{-\phi }g_{ab}.
\end{equation}

In this frame the 4-dimensional string field equations and the equations of
motion are given by (indices run $0\leq a,b,c\leq 3$ and $1\leq $ $\alpha
,\beta \leq 3$),

\begin{eqnarray}
R_{ab}-\frac 12g_{ab}R &=&\kappa ^2(^{(\phi )}T_{ab}+^{(H)}T_{ab}), \\
\nabla _a(e^{-2\phi }H^{abc}) &=&0, \\
\Box \phi +\frac 16e^{-2\phi }H_{abc}H^{abc} &=&0,
\end{eqnarray}
where $\kappa ^2=8\pi G$ is the 4-dimensional Einstein gravitational
coupling and 
\begin{eqnarray}
^{(\phi )}T_{ab} &\equiv &\frac 12(\phi _{,a}\phi _{,b}-\frac 12g_{ab}\phi
_{,c}\phi ^{,c}), \\
^{(H)}T_{ab} &\equiv &\frac 1{12}e^{-2\phi }(3H_{acd}H_b^{\;\;cd}-\frac
12g_{ab}H_{mlk}H^{mlk}).
\end{eqnarray}

The 3-geometries of the nine spatially homogeneous cosmological solutions of
these equations are defined by the Bianchi classification of homogeneous
spaces (with the exception of the Kantowski-Sachs universe, [9], which has a
four-dimensional group of motions but no three-dimensional subgroup).
In these Bianchi models (e.g. [13]) the spacelike hypersurfaces 
are invariant under
the group $G_{3}$ of isometries whose generators are 3 Killing vectors
$\xi_{\alpha}$. These hypersurfaces can be described by an invariant
vector basis $\{X_{\alpha}\}$ satisfying 
$${\cal L}_{\xi_{\beta}}X_{\alpha}=[\xi_{\beta},X_{\alpha}]=0$$
where ${\cal L}_{\xi_{\beta}}$ is the Lie derivative in the direction of
$\xi_{\beta}$.
The timelike direction $X_{0}$ is chosen to be orthogonal to the
invariant spacelike hypersurfaces obeying
$${\cal L}_{\xi_{\beta}}X_{0}=[\xi_{\beta},X_{0}]=0 .$$

Dual to  $\{X_{\alpha}\}$ is the basis of one-forms
$\{\omega^{\mu}\}$ satisfying
$$\omega^{\mu}=\frac{1}{2}
C^{\mu}_{\kappa\lambda}\omega^{\kappa}\wedge\omega^{\lambda} .$$

Spatial homogeneity is expressed by the following conditions on $\phi$,
$g$ and $H$
$${\cal L}_{\xi_{\alpha}}\phi=0 ,$$
$${\cal L}_{\xi_{\alpha}}g=0 ,$$
$${\cal L}_{\xi_{\alpha}}H=0\Rightarrow {\cal L}_{\xi_{\alpha}}(\ast
H)=0 .$$

The definition and properties of the Lie derivative imply that
${\cal L}_{\xi_{\alpha}}\phi=\xi_{\alpha}\phi=0$.
Expanding $\ast H$ in the invariant basis (that is,
$\ast H=V^{0}X_{0}+V^{\alpha}X_{\alpha}$), and using its properties,
implies then  $\xi_{\alpha}V^{0}=0$ and $\xi_{\alpha}V^{\beta}=0$.
The Killing vectors in the Bianchi models are spacelike and time
independent and this then implies that  $\phi$ and $H$ are functions 
of time only in the standard basis $\{dt,\omega^{\alpha}\}$.
Furthermore, the antisymmetric tensor potential $B$ where $H=dB$ will
be assumed to be a function of time only.

We would like to know the general algebraic form of the $H$ field  with
a time-dependent potential $B$ in these
models, determine which Bianchi universes are the most general, and discover
whether the assumption of spatial homogeneity reduces the number of
independent pieces of Cauchy data below the number needed to specify a
generic inhomogeneous solution of the field equations (4)-(8). This analysis
of the allowed components of the $H$-field is most economically performed
using differential forms.\\

\vspace{0.5cm}

\begin{center}
{\bf II.} {\bf THE ANTISYMMETRIC TENSOR FIELD AS A 2-FORM}
\end{center}

There are three equations determining the antisymmetric tensor field: the
definition of its field strength (for a closed bosonic string) 
\begin{eqnarray}
H=dB,
\end{eqnarray}
which implies the second equation$^{\ }$ 
\begin{eqnarray}
dH=0,
\end{eqnarray}
and there is the equation of motion, (5),

\begin{eqnarray}
d(*H)-2(d\phi )\wedge (*H)=0.
\end{eqnarray}

Spatially homogeneous models are described by choosing an orthonormal
tetrad, 
\begin{equation}
ds^2=\eta _{ab}\sigma ^a\sigma ^b,
\end{equation}
where $\eta _{ab}=diag(-1,1,1,1),$ and specifying the 1-forms ${\sigma ^a}$
[10,13] as

\begin{eqnarray}
\sigma ^0 &=&N(\Omega )d\Omega ,  \nonumber \\
\sigma ^\alpha  &=&e^{-\Omega }b_\beta ^\alpha \omega ^\beta .
\end{eqnarray}
Here, the $\omega ^\alpha $ obey the algebra 
\begin{eqnarray}
d\omega ^\alpha =\frac 12C_{\beta \gamma }^\alpha \omega ^\beta \wedge
\omega ^\gamma ,
\end{eqnarray}
where $C_{\beta \gamma }^\alpha $ are the structure constants of the
possible isometry groups which define the homogeneous 3-spaces, and the $%
b_\beta ^\alpha $ are symmetric matrices which depend only on the time
coordinate $\Omega $. Since $B$ is a 2-form, it can be decomposed as 
\begin{eqnarray}
&&  \nonumber \\
\ B &=&B_{0\alpha }\sigma ^0\wedge \sigma ^\alpha +B_{\alpha \beta }\sigma
^\alpha \wedge \sigma ^\beta =Q_{0\kappa }d\Omega \wedge \omega ^\kappa
+S_{\kappa \mu }\omega ^\kappa \wedge \omega ^\mu ,
\end{eqnarray}
where 
\begin{equation}
Q_{0\kappa }(\Omega )\equiv NB_{0\alpha }e^{-\Omega }b_\kappa ^\alpha ,
\end{equation}
\begin{equation}
S_{\kappa \mu }(\Omega )\equiv e^{-2\Omega }B_{\alpha \beta }b_\kappa
^\alpha b_\mu ^\beta .
\end{equation}
Hence, $H=dB$ is given by 
\begin{eqnarray}
H=(S_{\alpha \beta \mid \Omega }-\frac 12C_{\alpha \beta }^\kappa Q_{0\kappa
})d\Omega \wedge \omega ^\alpha \wedge \omega ^\beta +\frac 12S_{\kappa \mu
}C_{\alpha \beta }^{[\kappa }\omega ^{\mu ]}\wedge \omega ^\alpha \wedge
\omega ^\beta .
\end{eqnarray}
This expression can be analysed further if we introduce the Ellis-MacCallum
[11] decomposition of the structure constants into the matrix $m_{\alpha
\beta }\ $and the vector $a_\beta ,$

\begin{equation}
C_{\alpha \beta }^\gamma =\epsilon _{\alpha \beta \mu }m^{\mu \gamma
}+\delta _\beta ^\gamma a_\alpha -\delta _\alpha ^\gamma a_\beta ,
\end{equation}
so (18) becomes

\begin{eqnarray}
H=(S_{\alpha \beta \mid \Omega }-\frac 12C_{\alpha \beta }^\kappa Q_{0\kappa
})d\Omega \wedge \omega ^\alpha \wedge \omega ^\beta +2a_\alpha S_{\kappa
\mu }\omega ^\mu \wedge \omega ^\alpha \wedge \omega ^\kappa .
\end{eqnarray}
The structure constants satisfy a Jacobi identity which leaves $C_{\alpha
\beta }^\gamma $ with a maximum of 6 independent components. Since the
lagrangian is invariant under the gauge transformation $B_{ab}\rightarrow
B_{ab}+\partial _{[a}\Lambda _{b]},$ we can always choose $\Lambda $ such
that $Q_{0\kappa }=-\partial _{[0}\Lambda _{\kappa ]}=-\partial _0\Lambda
_\kappa ,$ and set $Q_{0\kappa }$ to be zero.

The nine Bianchi-type universes fall into two classes, A and B,
distinguished by whether the constant $a$ is zero or non-zero respectively
[11]. From (20) we see that $H$ has no purely spatial components in Class A
models.

$H$ is also given by

\begin{eqnarray}
H &=&H_{abc}\sigma ^a\wedge \sigma ^b\wedge \sigma ^c  \nonumber \\
&=&H_{0\alpha \beta }\sigma ^0\wedge \sigma ^\alpha \wedge \sigma ^\beta
+H_{\alpha \beta \gamma }\sigma ^\alpha \wedge \sigma ^\beta \wedge \sigma
^\gamma  \nonumber \\
&=&X_{0\kappa \lambda }d\Omega \wedge \omega ^\kappa \wedge \omega ^\lambda
+Y_{\kappa \lambda \mu }\omega ^\kappa \wedge \omega ^\lambda \wedge \omega
^\mu ,
\end{eqnarray}
where 
\begin{equation}
X_{0\kappa \lambda }=X_{0\kappa \lambda }(\Omega )\equiv Ne^{-2\Omega
}b_\kappa ^\alpha b_\lambda ^\beta H_{0\alpha \beta },
\end{equation}
\begin{equation}
Y_{\kappa \lambda \mu }=Y_{\kappa \lambda \mu }(\Omega )\equiv e^{-3\Omega
}b_\kappa ^\alpha b_\lambda ^\beta b_\mu ^\gamma H_{\alpha \beta \gamma }.
\end{equation}
Therefore $dH=0$ implies

\begin{eqnarray}
Y_{\kappa \lambda \mu \mid \Omega }d\Omega \wedge \omega ^\kappa \wedge
\omega ^\lambda \wedge \omega ^\mu +\frac 12X_{0\kappa \lambda }(C_{\alpha
\beta }^\kappa \omega ^\lambda -C_{\alpha \beta }^\lambda \omega ^\kappa
)\wedge \omega ^\alpha \wedge \omega ^\beta \wedge d\Omega =0.
\end{eqnarray}

Using the expression (19) for the structure constants, and noting that the
3-dimensional Levi-Civita symbol is defined by $\epsilon =\sqrt{\det
g_{\alpha \beta }\ }\omega ^1\wedge \omega ^2\wedge \omega ^3,$ eq. (24)
becomes

\begin{eqnarray}
(Y_{\kappa \lambda \mu \mid \Omega }-2X_{0\mu \kappa }a_\lambda )d\Omega
\wedge \omega ^\kappa \wedge \omega ^\lambda \wedge \omega ^\mu =0.
\end{eqnarray}
The dual, $*\lambda ,$ of an n-dimensional p-form $\lambda $ is defined by
the Levi-Civita symbol as [12] 
\[
\ast \lambda _{b_1...b_{n-p}}=\frac 1{p!}\lambda ^{a_1...a_p}\epsilon
_{a_1...a_pb_1...b_{n-p}}. 
\]
Hence, $*H$ is a 1-form given by 
\begin{eqnarray}
&&  \nonumber \\
\ast H &=&\frac 16H^{bcd}\epsilon _{bcda}\sigma ^a=Ud\Omega +V_\alpha \omega
^\alpha ,
\end{eqnarray}
where 
\begin{equation}
U\equiv U(\Omega )\equiv \frac 16H^{\alpha \beta \gamma }\epsilon _{\alpha
\beta \gamma 0}N,
\end{equation}

\begin{equation}
V_\alpha \equiv V_\alpha (\Omega )\equiv \frac 16H^{abc}\epsilon _{abc\kappa
}b_\alpha ^\kappa e^{-\Omega },
\end{equation}
and so$\ $

\begin{equation}
d*H=V_{a\mid \Omega }d\Omega \wedge \omega ^\alpha +\frac 12V_\alpha
C_{\beta \gamma }^\alpha \omega ^\beta \wedge \omega ^\gamma .
\end{equation}

Since $\phi=\phi(\Omega)$, we have $d\phi =\phi _{\mid \Omega }d\Omega $
and equation (11) reads

\begin{eqnarray}
(V_{\alpha \mid \Omega }-2\phi _{\mid \Omega }V_\alpha )d\Omega \wedge
\omega ^\alpha +\frac 12V_\alpha C_{\beta \gamma }^\alpha \omega ^\beta
\wedge \omega ^\gamma =0;
\end{eqnarray}
hence, 
\begin{eqnarray}
V_{\alpha \mid \Omega }-2\phi _{\mid \Omega }V_\alpha &=&0, \\
\frac 12V_\alpha C_{[\beta \gamma ]}^\alpha &=&0.
\end{eqnarray}

Notice that the constraint (32) is preserved in time. Contracting (31) with $%
C_{[\beta \gamma ]}^\alpha $ gives $(V_\alpha C_{[\beta \gamma ]}^\alpha
)_{\mid \Omega }=0$ so that if (32) is satisfied at one time it holds at all
times. Eqn. (32) implies 
\begin{equation}
\epsilon ^{\beta \gamma \delta }V_\alpha C_{\beta \gamma }^\alpha =0,
\end{equation}
which can be rewritten as 
\begin{eqnarray}
V_\alpha (m^{\delta \alpha }+a_\beta \epsilon ^{\beta \alpha \delta })=0,
\end{eqnarray}
and so, by (20), we have 
\begin{equation}
H=\ X_{0\kappa \lambda }d\Omega \wedge \omega ^\kappa \wedge \omega ^\lambda
+Y_{\kappa \lambda \mu }\ \omega ^\mu \wedge \omega ^\alpha \wedge \omega
^\kappa ,
\end{equation}
with 
\begin{equation}
X_{0\kappa \lambda }=S_{[\kappa \lambda ]\mid \Omega }-\frac 12C_{[\kappa
\lambda ]}^\nu Q_{0\nu }{\;\;\rm and\;\;}Y_{\kappa \lambda \mu }=2a_{[\alpha
}S_{\kappa \mu ]}.
\end{equation}
Eqn. (25) implies 
\begin{eqnarray}
C_{[\mu \kappa }^\alpha a_{\lambda ]}Q_{0\alpha }=0,
\end{eqnarray}
and (31) can be integrated to give

\begin{eqnarray}
V_\alpha =e^{2\phi }K_\alpha ,
\end{eqnarray}
where $K_\alpha $ is a constant spatial 3-vector of integration.

Since $*(*H)=H,$ we have 
\begin{eqnarray}
X^{0\alpha \beta } &=&-\epsilon ^{0\alpha \beta \gamma }V_\gamma  \nonumber
\\
\ &=&\epsilon ^{0\alpha \beta \gamma }e^{2\phi }K_\gamma ,
\end{eqnarray}
where the minus sign has been absorbed into the constant spatial 3-vector $%
K_\gamma $.\\

\begin{center}
TABLE 1\\
\end{center}

Table 1 displays the restrictions on the spatial components of $*H$ imposed
by the constraint equation (34) for the different Bianchi types [11,13],
together with the components of the homogeneous antisymmetric tensor field
strength $H$ in the standard basis $\{d\Omega ,\omega ^\alpha \}$ which are
given by eqn. (20). Note that in Class A, eqn. (23) implies $Y_{123}=0,$ and
the contravariant components of $Y_{\alpha \beta \gamma }$ are obtained by
raising the indices using $g_{ab}$ given by 
\begin{equation}
g_{00}=-N^2(\Omega )
\end{equation}
\begin{equation}
g_{\alpha \beta }=e^{-2\Omega }\sum_\gamma b_\alpha ^\gamma b_\beta ^\gamma .
\end{equation}
In Class B, eqn.(23) implies that $Y_{123}=2a_{[2}S_{31]}=2a_3S_{12}.$ The
matrix ${\bf \alpha }$ which specifies the Ellis-MacCallum symbol ${\bf m}%
=m_{\alpha \beta }$ is defined by, [11], the matrix

\begin{equation}
{\bf \alpha }=\left( 
\begin{tabular}{lcr}
0 & 1 & 0 \\ 
1 & 0 & 0 \\ 
0 & 0 & 0
\end{tabular}
\right)
\end{equation}

\vspace{0.5 cm}

\begin{center}
{\bf III. COUNTING DEGREES OF FREEDOM}
\end{center}

Consider first the question of how many independently arbitrary spatial
functions are required to specify generic initial data for the system of
string field equations (4)-(8). In a synchronous frame we require 6 $%
g_{\alpha \beta }$, 6 $\dot g_{\alpha \beta },$ 3 components of the $H\ $%
field, together with values of $\phi $ and $\dot \phi .$ This amounts to 17
functions, but we can remove 4 by using the coordinate covariance of the
theory, another 4 by using the $R_{0a}$ constraint equations, and another 1
by using the $\phi $ equation, (6). This leaves 8 independent functions of
three spatial variables to specify a general solution of the field equations
(4)-(8). If special symmetries are assumed for the solutions of the field
equations then some of the metric components and their time derivatives may
be absent but some of the algebraic $R_{0a}$ constraints may be identically
satisfied. As a result, the number of functions characterising the most
general solution compatible with some symmetry may be specified by fewer
functions (or by lower-dimensional functions) than the general solution.

Spatially homogeneous cosmological models will be determined by some number
of independently arbitrary constants rather than spatial functions. If
spatially homogeneous string cosmologies are representative of the most
general inhomogeneous string cosmologies then it is necessary (although not
necessarily sufficient) that they be characterised by 8 independent
arbitrary constants. When the $H$ field vanishes in eqns. (4)-(8), so they
reduce Einstein's equations for a free scalar field, the number of arbitrary
functions is required to characterise the general inhomogeneous solution
equals the number of constants required for the general homogeneous
solution. This equivalence also holds for Einstein's equations with a
perfect fluid (or in vacuum), where 8 (or 4) functions specify a general
inhomogeneous solution and 8 (or 4) constants specify Bianchi types VI$_h,$
VII$_h$, VIII, and IX, [15,16]. We shall now investigate the degree of
generality of the different Bianchi type solutions of the string field
equations when the $H$ field is present.

In order to determine how many free parameters are allowed in the different
Bianchi models, consider the field equations, (4), for spatially homogeneous
universes in the standard basis $\{d\Omega ,\omega ^\alpha \}.$ The
components of the Ricci tensor are given by [14] 
\begin{eqnarray}
R_{00} &=&-\dot \theta -\theta _{\alpha \beta }\theta ^{\alpha \beta }, \\
R_{0\alpha } &=&3a_\gamma \theta _\alpha ^\gamma -a_\alpha \theta +\epsilon
_{\gamma \alpha \tau }m^{\tau \beta }\theta _\beta ^\gamma , \\
R_{\alpha \beta } &=&\dot \theta _{\alpha \beta }+\theta \theta _{\alpha
\beta }-2\theta _{\alpha \gamma }\theta _\beta ^\gamma +\Gamma _{\lambda
\gamma }^\gamma \Gamma _{\alpha \beta }^\lambda -\Gamma _{\lambda \beta
}^\gamma \Gamma _{\alpha \gamma }^\lambda +C_{\gamma \beta }^\kappa \Gamma
_{\alpha \kappa }^\gamma ,
\end{eqnarray}
where $\theta _{\alpha \beta }=\frac 12g_{\alpha \beta \mid \Omega }$ , $%
\theta \equiv \theta _\alpha ^\alpha $ and the Ellis-MacCallum
parametrization, (19), has been used to express the spatial curvature terms
in (44) and (45).

The string field equations give 10 equations for the 6 components of the
symmetric metric $g_{\alpha \beta },$ so there are at most 4 constraint
equations. The initial data for $g_{\alpha \beta }$ consist of 12
independent constants: 6 $g_{\alpha \beta }$ and 6 $\dot g_{\alpha \beta }$.
These are reduced by $(9-p+1)$ due to the fact that there are $9-p+1$
parameters of triad freedom to put the group structure constants into their
canonical Ellis-MacCallum form [14]. The parameter $p$ is the number of
independent group structure constants and $0\leq p\leq 6$. Their values are
given below, and in Table 1, for each Bianchi group type. The number of
independent constants is reduced by a further $\ 4-r$ due to the constraint
equations, where $r$ counts the number of field equations satisfied
identically. Hence, a total of $12-(9-p+1)-(4-r)=p+r-2$ independent
constants specify the general solution to equations (4)-(8) for spatially
homogeneous universes. To calculate $r$ we must check if any of the field
equations are identically satisfied due to a particular choice of the group
structure parameters $a_\beta $ and $m_{\alpha \beta }$. From eqn. (7) it is
clear that the dilaton's contribution to the $R_{0\alpha }$ equations
vanishes identically. The contribution by the $H$ field is determined by $%
H_{0cd}H_\alpha ^{\;\;cd},$ but we know from (35)-(36) that $H_{0cd}H_\alpha
^{\;\;cd}=-X^{0\beta \gamma }Y_{\alpha \beta \gamma },$ hence $R_{0\alpha }=0
$ for all Class A models.

The equations of motion for $H$ and the constraints they impose have been
discussed in section II; Table 1 gives the number of free parameters, $3-u,\
\ $to specify initial data for $H$ for each group type. The initial
conditions for the dilaton $\phi $ require 2 further independent constants: $%
\phi $ and $\dot \phi ,$ while eqn.(6) determines the dynamics of $\phi $.
Therefore the general spatially homogeneous solution(s) to equations (4)-(8)
contain 
\begin{eqnarray}
{\cal N}\equiv 3+p+r-u
\end{eqnarray}
independent arbitrary constants. Using the constraint equations (34) and
(44) we can evaluate $p,r,u,$ and ${\cal N}$(Bianchi type) explicitly as
follows (the values of these parameters are summarised in Table 1).

\begin{center}
{\bf A. Class A models}\\
\end{center}

{\it Bianchi I}: $R_{0\alpha }=0,$ hence $r=3$, $p=0$, $u=0$ and ${\cal N}%
(I)=6$

{\it Bianchi II}: $R_{01}=0,R_{02}=-\theta _1^3,R_{03}=\theta _1^2,$ hence $%
r=1$, $p=3$, $u=1,$ and ${\cal N}(II)=6$

{\it Bianchi VI}$_{-1}$:$\ R_{01}=-\theta _1^3$, $R_{02}=\theta
_2^3,R_{03}=\theta _1^1-\theta _2^2,$ hence $r=0$, $p=5$, $u=2,$ and ${\cal N%
}(VI_{-1})=6$

{\it Bianchi VII}$_0$: $R_{01}=-\theta _2^3$, $R_{02}=\theta _1^3$, $%
R_{03}=\theta _2^1-\theta _{1\;\;\;}^2,$ hence $r=0$, $p=5$, $u=2,$ and $%
{\cal N}(VII_0)=6$

{\it Bianchi VIII}: $R_{01}=\theta _2^3-\theta _3^2,R_{02}=\theta
_1^3+\theta _3^1,R_{03}=-\theta _1^2-\theta _2^1,$ hence $r=0$, $p=6$, $u=3$
and ${\cal N}(VIII)=6$

{\it Bianchi IX}: $R_{01}=\theta _2^3-\theta _3^2,$ $R_{02}=\theta
_3^1-\theta _{1\;\;\;}^3,R_{03}=\theta _1^2-\theta _2^1,\ $hence $r=0$, $%
p=6$, $u=3,$ and ${\cal N}(IX)=6.$

Hence, all Class A models are equally general according to the
parameter-counting criterion.

\begin{center}
{\bf B. Class B models}\\
\end{center}

{\it Bianchi III}: $R_{01}=-2\theta _1^3,R_{02}=-\theta _2^3,R_{03}=\theta
_1^1-\theta _3^3,$ hence $r=0$, $p=5$, $u=0,$ and ${\cal N}(III)=8$

{\it Bianchi IV}: $R_{01}=-3\theta _1^3,R_{02}=-3\theta _2^3-\theta
_1^3,R_{03}=-\theta _3^3+\theta +\theta _1^2,$ hence $r=0$, $p=5$, $u=1,$
and ${\cal N}(IV)=7$

{\it Bianchi V}: $R_{01}=-3\theta _1^3,R_{02}=-3\theta _2^3,R_{03}=-\theta
-3\theta _3^3,$ hence $r=0$, $p=3$, $u=1,$ and ${\cal N}(V)=5$

{\it Bianchi VI}$_{h\neq -1}$: $R_{01}=-(h+2)\theta
_1^3,R_{02}=-(2h+1)\theta _2^3,R_{03}=\theta _1^1+h\theta _2^2-(h+1)\theta
_3^3.$ For special choices of $h,$ either $R_{01}$ or $R_{02}$ can be made
to vanish identically. The two choices are either $h=-2$ or $h=-\frac 12,$
[15], so that $r(h=-2)=r(h=-1/2)=1$. Therefore, we have three cases,

\begin{tabular}{lcr}
(i) $h=0:r=0,\;p=5,u=0,$ and ${\cal N}(VI_0)=8,$ &  &  \\ 
(ii) $h\neq 0$, $h\neq -2$ and $h\neq -\frac 12:r=0,\;p=5,$ $u=1,$ and $%
{\cal N}(VI_h)=7,$ &  &  \\ 
(iii) $h=-2$ or $h=-\frac 12:r=1,\;p=5,\;u=1,$ and ${\cal N}(VI_{h\neq
-2,-\frac 12})=8.$ &  & 
\end{tabular}

{\it Bianchi VII}$_{h\neq 0}$: $R_{01}=-h\theta _1^3-\theta _2^3,$ $%
R_{02}=\theta _1^3-2h\theta _2^3,R_{03}=\theta _2^1-\theta _1^2+h\theta
_2^2-h\theta _3^3.$ Since $h$ is a real parameter there are no exceptional
cases; hence $r=0$, $p=5$, $u=1,$ and ${\cal N}(VII_h)=7.$\\

This analysis of Class B models indicates that they are described by more
free parameters than those of Class A and the different Class B models are
not equally general like those of Class A. Thus the most general spatially
homogeneous solutions of the string equations (4)-(8) are those of Bianchi
types III and VI$_{h=0,-\frac 12,-2}.$ These cosmological models contain the
maximum number of 8 free parameters. These results can be compared with the
study of homogeneous pure magnetic or pure electric fields in
general-relativistic Bianchi universes carried out by Hughston and Jacobs
[10] and Ruban [17] where a similar phenomenon occurs. For a homogeneous
magnetic field the most general solution is found to be Bianchi type III.
Contrary to what was found for the antisymmetric tensor field strength, the
exceptional purely magnetic universes of Bianchi type VI do not contain as
many free parameters as the Bianchi III universe. This might be related to
the fact that $H$ is a 3-form and the homogeneous space is three
dimensional, which implies that $dH$ does not provide additional constraints
on the purely spatial components of $H,$ whereas in the case of a Maxwell
2-form, $f,$ the differential, $df,$ does give additional constraints on the
purely spatial components involving the group structure constants via the
Maxwell equations $df=0=d*f$. However, the number of free parameters in
Bianchi models in the cases of the homogeneous pure magnetic fields and
homogeneous antisymmetric tensor field strength possess common features. In
both cases the generality of the solutions is the same in Bianchi types IV,
VI$_{h\neq 0,-\frac 12,-2}$ and VII$_{h\neq 0},$and the least general model
is Bianchi type V.

The hierarchy of generality in the string cosmologies has several
interesting features when compared with the situation of vacuum and
perfect-fluid universes in general relativity. The most general category of
8-parameter models (types III and VI$_{h=0,-\frac 12,-2})$ does not contain
closed universes (ie type IX) as in general relativity, nor does it contain
any types which contain isotropic universes as particular cases (ie types I,
V, VII$_0,$VII$_h$ or IX). Isotropy is not an open property of homogeneous
initial data space. This is related to the fact that the $H$ field is an
anisotropic stress: the isotropic limit cannot be obtained with a non-zero $%
H $ field. This means that the isotropic Friedmann universes appear to be
even less representative of the general behaviour of cosmological models in
string theory than they are in general relativity. However, a similar
situation can arise in general relativity when anisotropic stresses are
included.

\vspace{0.5 cm}

\begin{center}
{\bf IV. CONCLUSIONS}
\end{center}

The equations that determine the antisymmetric tensor field in low-energy
effective string theory have been investigated in spatially homogeneous
Bianchi-type universes. It is found that the homogeneous 3-form $H$ with
a homogeneous potential can have
at most three nonvanishing components. The number of allowed components were
fully classified in Table 1. In Bianchi Class A models the field strength $H$
has no purely spatial components in the standard basis. Bianchi types VIII
and IX allow only a time-independent, antisymmetric tensor field, $B_{\mu
\nu },$ which implies a vanishing field strength $H$. In the case of Bianchi
IX this can be understood in geometrical terms. Each of the Bianchi models
corresponds to a group of motions or isometries of spatial hypersurfaces. In
the case of Bianchi IX this group is isomorphic to SO(3,I$\!$R), which is
isomorphic to the three-dimensional rotation group. Since the dual of the
antisymmetric tensor field strength, $H,$ is a vector, one of the spatial
directions is picked out and this is incompatible with the rotational
invariance.

In comparison with  Batakis'  findings on the possible configurations of
the $H$-field (not necessarily derived from a homogeneous potential) 
in diagonal Bianchi models [6] [5] the cases $\chi(d \rightarrow)$ and
$\chi(d \nearrow)$ are recovered [18]
if Einstein's equations for the diagonal Bianchi $IV$ and $VII$ models
are taken into account (primarily the $R_{12}$ constraint equation
in the
orthonormal frame which implies a solution which is singular everywhere).
For the $\chi(d\rightarrow)$ case one must bear in mind that
for  $Y_{123}=0$ equation (25) implies $X_{012}=0$.
Since we started with a purely time-dependent potential the case $\chi(d\uparrow)$
is only partially recovered.
However, in the other two cases the generality is not restricted by
assuming a purely time-dependent potential.   

Eight independent functions of three spatial variables were found to be
required to characterise a general inhomogeneous solutions of the string
field equations. This was compared with the number required to specify each
homogeneous Bianchi type solution. It was found that the most general
homogeneous solutions are of Bianchi types III and VI$_{h=0,-\frac 12,-2}$,
and contain 8 independent constants. This situation contrasts with that for
spatially homogeneous vacuum and perfect-fluid universes in general
relativity and for degenerate string cosmologies with $\phi \neq 0$ and $H=0$%
. In these cases, the most general universes are of Bianchi types VI$_h,$ VII%
$_h$, VIII, and IX. When $H\neq 0$ we find a change in the relative degrees
of generality that is analogous to that found in the case of spatially
homogeneous general relativistic universes containing pure magnetic fields.%

\vspace{0.5 cm}

{\bf Acknowledgements }The authors would like to thank A. Lahiri and M.
Dabrowski for discussions. JDB was supported by a PPARC Senior Fellowship
and KEK was supported by the German National Scholarship Foundation.\\ 

\vspace{0.5 cm}

{\bf \ References}\\

[1] M. B. Green, J. H. Schwarz, E. Witten {\it Superstring Theory,}{\sl \ }%
Vol. I (CUP: Cambridge, 1987); E. S. Fradkin, \& A. A. Tseytlin, Nucl. Phys. 
{\bf B 261,} 1 (1985); C. G. Callan, E. J. Martinec \& M. J. Perry, Nucl.
Phys. {\bf B} {\bf 262}, 593 (1985); C. Lovelace, Nucl. Phys.{\bf \ B} {\bf %
273}, 413 (1985).

[2] E. J. Copeland, A. Lahiri, \& D. Wands, Phys. Rev. D {\bf 50}, 4868
(1994).

[3] E. J. Copeland, A. Lahiri, \& D. Wands, Phys. Rev. D {\bf 51,} 1569
(1995)

[4] M. Gasperini, J. Maharana \& G. Veneziano, Phys. Lett. B {\bf 272}, 277
(1991); J. Garc\'{i}a-Bellido, M. Quir\'{o}s, Nucl. Phys. B{\bf 368} 463
(1992); J. Garc\'{i}a-Bellido, M. Quir\'{o}s, Nucl. Phys. B {\bf 385} 558 (1992);
M. Gasperini, \& G. Veneziano, Phys. Lett. B {\bf 277}, 256 (1992);
M. Gasperini, R. Ricci \& G. Veneziano, Phys. Lett. B {\bf 319}, 438 (1993);
M. Gasperini \& R. Ricci, Class. Quantum Grav. {\bf 12}, 677 (1995). A. A.
Saaryan, Astrophysics {\bf 38}, 164 (1995).

[5] N. A. Batakis \& A. A. Kehagias, Nucl. Phys. B {\bf 449,} 248 (1995); N.
A. Batakis, Phys. Lett. B {\bf 353,} 450 (1995).

[6] N. A. Batakis, Phys. Lett. B {\bf 353,} 39 (1995).

[7] L. Bianchi, Mem. Soc. It. della Sc. (dei XL) {\bf 11}, 267 (1897); {\it %
Lezioni sulla teoria dei gruppi continui finiti di transformazioni},
(Spoerri: Pisa, 1918).

[8] A. H. Taub, Ann. Math. {\bf 53}, 472 (1951).

[9] R. Kantowski \& R. K. Sachs, J. Math. Phys. {\bf 7}, 443 (1966); A. S.
Kompanyeets \& A .S. Chernov, Sov. Phys. JETP {\bf 20}, 1303 (1964); C. B.
Collins, J. Math. Phys. {\bf 18}, 2116 (1977). Exact string cosmological
solutions have been found for Kantowski-Sachs models by J. D. Barrow and M.
Dabrowski, Phys. Rev. D {\bf 000} 000 (1996).

[10] L. P. Hughston \& K. C. Jacobs, Astrophys. J. {\bf 160,} 147 (1969).

[11] G. F. R. Ellis \& M. A. H. MacCallum, Comm. Math. Phys. {\bf 12,} 108
(1969) and {\bf 19,} 31 (1970).

[12] See e.g. R. M. Wald, {\it General Relativity }(U. Chicago P.: Chicago,
1984).

[13] M. P. Ryan \& L. C. Shepley, {\it Homogeneous Relativistic Cosmologies, 
}(Princeton U. P.; Princeton, 1975).

[14] M. A. H. MacCallum, in {\it General Relativity: an Einstein Centenary, }%
eds. S. W. Hawking \& W. Israel (CUP: Cambridge, 1979); S. T. C. Siklos, in 
{\it Relativistic Astrophysics and Cosmology}, eds. X. Fustero \& E.
Verdaguer, (World Scientific: Singapore, 1984), pp 201-248; S. T. C. Siklos,
Class. Q. Grav. {\bf 13}, 1931 (1996); J. D. Barrow \& D. H. Sonoda, Phys.
Rep. {\bf 139},1 (1986); J.D. Barrow, in {\it The Physical Universe: The
Interface Between Cosmology, Astrophysics and Particle Physics}, eds. J. D.
Barrow, A.B. Henriques, M.T.V.T. Lago \& M.S. Longair, Springer LNP 383,
(Springer: NY,1991), pp1-20.

[15] L. Landau \& E. M. Lifshitz, {\it The Classical Theory of Fields,} 4th
edn., (Pergamon: Oxford, 1975).

[16] $h=-\frac 19$ is not the special value because Ryan and Shepley, in
ref. [13], use an off-diagonal parametrization for $m_{\alpha \beta }$ in
Bianchi VI$_{h\neq -1}$; $h=-\frac 19$ is obtained as special value if one
chooses the diagonal parametrization.

[17] V. A. Ruban, Sov. Astron. {\bf 26}, 632 (1983) and {\bf 29}, 5 (1985).

[18] $\chi(d\;\;\;)$ is a parameter with value I,II,...,IX denoting
the Bianchi classification, $d$ stands for diagonal and an arrow
indicates the orientation of $\ast H=\ast H_{0}(t)dt+\ast H_{\alpha}(t)
\omega^{\alpha}$. So that  in a $\chi(d\uparrow)$ model $\ast H$ has
only a timelike component, in a $\chi(d\rightarrow)$ model only spatial
components while in a $\chi(d \nearrow)$ model both components appear.


\begin{table}
\begin{center}
\begin{tabular}{|l||c|c||c|c|c||c|c|c|c|}
\hline
Bianchi Type & $a_{\alpha}$ & {\bf m} & $V_{\alpha}$ & $X^{0\alpha\beta}$ &
$Y_{123}$ & $p$ & $ r$ & $u$ & ${\cal N}$\\
\hline\hline
 & & 
&            & $X^{012}=e^{2\phi}K_{3}$ &
& & & &  \\                    

{\bf I} & 0 & 0 
& $V_{\alpha}$ arb & $X^{013}=-e^{2\phi}K_{2}$ & 0
& 0 & 3 & 0 & 6\\

 & &
&            & $X^{023}=e^{2\phi}K_{1}$ &  
& & & &  \\
\hline            
 & &
& $V_{1}=0$ & $X^{012}=e^{2\phi}K_{3}$ &
& & & &  \\

{\bf II} & 0 & diag(1,0,0) 
& $V_{2}$ arb & $X^{013}=-e^{2\phi}K_{2}$ & 0
& 3 & 1 & 1 & 6\\

 & &
& $V_{3}$ arb & $X^{023}=0$ &
& & & &  \\
\hline

 & &
& $V_{1}=0$ & $X^{012}=e^{2\phi}K_{3}$ &
& & & &  \\

{\bf VI$_{-1}$} & 0 & {\bf m}=$-{\bf\alpha}$ 
&$V_{2}=0$ & $X^{013}=0$ & 0
& 5 & 0 & 2 & 6\\

 & &
& $V_{3}$ arb & $X^{023}=0$ &
& & & &  \\
\hline

 & &
& $V_{1}=0$ & $X^{012}=e^{2\phi}K_{3}$ &
& & & &  \\

{\bf VII$_{0}$} & 0 & diag$(-1,-1,0)$ 
&$V_{2}=0$ & $X^{013}=0$ & 0
& 5 & 0 & 2 & 6\\

 & &
& $V_{3}$ arb & $X^{023}=0$ &
& & & &  \\
\hline

 & &
& $V_{1}=0$ & $X^{012}=0$ &
& & & &  \\

{\bf VIII} & 0 & diag$(-1,1,1)$
&$V_{2}=0$ & $X^{013}=0$ & 0
& 6 & 0 & 3 & 6\\

 & &
& $V_{3}$=0 &  $X^{023}=0$ &
& & & &  \\
\hline
    
 & &
& $V_{1}=0$ & $X^{012}=0$ &   
& & & &  \\

{\bf IX} & 0 & diag$(1,1,1)$
&$V_{2}=0$ & $X^{013}=0$ & 0
& 6 & 0 & 3 & 6\\

 & &
& $V_{3}$=0 & $X^{023}=0$ &
& & & &  \\  
\hline\hline

 & &
& $V_{1}=0$ & $X^{012}=e^{2\phi}K_{3}$ &
& & & &  \\

{\bf III} & $-\frac{1}{2}\delta^{\alpha}_{3}$ & $-\frac{1}{2}{\bf\alpha}$
& $V_{2}$ arb & $X^{013}=-e^{2\phi}K_{2}$ & $-S_{12}$
& 5 & 0 & 0 & 8\\

 & &
& $V_{3}$ arb & $X^{023}=0$ &
& & & &  \\
\hline

 & &
& $V_{1}=0$ & $X^{012}=e^{2\phi}K_{3}$ &
& & & &  \\

{\bf IV} & $-\delta^{\alpha}_{3}$ & diag(1,0,0)
& $V_{2}$=0 & $X^{013}=0$ & $-2S_{12}$
& 5 & 0 & 1 & 7\\

 & &
& $V_{3}$ arb & $X^{023}=0$ &
& & & &  \\
\hline

 & &
& $V_{1}=0$ & $X^{012}=e^{2\phi}K_{3}$ &
& & & &  \\

{\bf V} & $-\delta^{\alpha}_{3}$ & 0
& $V_{2}$=0 & $X^{013}=0$ & $-2S_{12}$
& 3 & 0 & 1 & 5\\

 & &
& $V_{3}$ arb & $X^{023}=0$ &
& & & &  \\   
\hline

$\;\;\; h=0$ & &
& $V_{1}=0$ & $X^{012}=e^{2\phi}K_{3}$ &
&5 &0 &0 &8   \\

{\bf VI$_{h\neq -1}$} & $-\frac{1}{2}(h+1)\delta^{\alpha}_{3}$ & $
-\frac{1}{2}(h-1)\alpha$ 
& $V_{2}h$=0 &   $ X^{013}_{h=0}=-e^{2\phi}K_{2}$ 
& $-(h+1)S_{12}$
&  & &  &  \\

$\;\;\; h\neq 0,-\frac{1}{2}, -2$ & &
& &$ X^{013}_{h\neq 0}$ = 0 &
& 5 & 0& 1 & 7 \\

$\;\;\; h=-\frac{1}{2}, -2$ & &
& $V_{3}$ arb & $X^{023}=0$ &
& 5 & 1 & 1 & 8 \\
\hline 

 & &
& $V_{1}=0$ & $X^{012}=e^{2\phi}K_{3}$ &
& & & &  \\

{\bf VII$_{h\neq 0}$} & $-\frac{h}{2}\delta^{\alpha}_{3}$ & 
diag$(-1,-1,0)$ 
& $V_{2}=0$ & $X^{013}=0 $ & $-\frac{h}{2}S_{12}$
& 5 & 0 & 1 & 7\\

 & &$ \;\;\;+ \frac{h}{2}{\bf\alpha}$
& $V_{3}$ arb & $X^{023}=0$ &
&  & &  &  \\

&&&&&&&&&\\
\hline

\end{tabular}
\end{center}
\caption{Summary of the possible components of the homogeneous antisymmetric tensor field strength 
and degrees of freedom. The different variables are explained in the text. }
\end{table}

\end{document}